\documentstyle[sprocl,psfig,rotate]{article}

\def\be{\begin{equation}}
\def\ee{\end{equation}}
\def\bea{\begin{eqnarray}}
\def\eea{\end{eqnarray}}

\begin{document}

\renewcommand{\textfraction}{0.}
\renewcommand{\topfraction}{1.}

\title{PHOTOABSORPTION ON NUCLEI}

\author{M. EFFENBERGER AND U. MOSEL}

\address{Institut f\"ur Theoretische Physik, Universit\"at Giessen\\
Heinrich-Buff-Ring 16, D-35392 Giessen}


\maketitle\abstracts{
We calculate the total photoabsorption cross section and cross sections for 
inclusive pion and eta photoproduction in nuclei 
in the energy range from 300 MeV to 1 GeV 
within the framework of
a semi-classical BUU transport model. Besides medium mo\-di\-fi\-ca\-tions
like Fermi motion and Pauli blocking we focus on the collision
broadening of the involved resonances. The resonance contributions
to the elementary cross section are fixed by fits to partial  
wave amplitudes of pion photoproduction. The cross sections for
$NR \to NN$, needed for the calculation of collision broadening,
are obtained by detailed balance from a fit to $NN \to NN\pi$ cross
sections. We show that a reasonable collision broadening is not
able to explain the experimentally observed disappearance of the
$D_{13}$(1520)-resonance in the photoabsorption cross section on nuclei.}

\section{Introduction}
The total nuclear photoabsorption cross section in the
first and second nucleon resonance region which was recently
measured in Mainz \cite{frommhold} and Frascati \cite{bianchi3,bianchi,bianchi2} 
shows clear medium modifications compared to the elementary 
cross section on the proton and deuteron \cite{daphne96,armstrongproton,armstrongdeut}. One observes a strong broadening of the $\Delta$-peak and
the disappearance of the higher resonances $D_{13}$ and $F_{15}$ while 
the total cross section per nucleon is almost independent of the mass
of the nucleus.
\par We present now a consistent 
calculation of the photon-nucleus reaction over the whole energy range
from 300 MeV to 1 GeV within
the framework of a semi-classical BUU transport model \cite{ber84,cass90}
which has been very successfully applied to the description of
heavy-ion collisions up to bombarding energies of 2 GeV/A \cite{wolf,teis} and
pion-nucleus reactions \cite{engel93}. 
Our calculation is based on the assumption that the total photonuclear
cross section is the incoherent sum of contributions from all nucleons
where we neglect possible shadowing effects.
Besides more or less trivial medium modifications
like Fermi motion and Pauli blocking we investigate the effect of
collision broadening for the involved resonances. This may lead to a 
better understanding of the behaviour of nucleon resonances in nuclear
matter.
\par Pion photoproduction on nuclei is determined both by the elementary 
$(\gamma,\pi)$ process on the nucleon as well as by final state $\pi$-N 
interactions whereas photoabsorption is dominated by the former reaction.
A detailed investigation of $(\gamma,\pi)$ on nuclei could thus help to
separate these two effects and to identify true in-medium effects on the
primary production process.
\par In section \ref{model} we start with a brief presentation of the used
BUU model. 
The results for the collision widths of the nucleon resonances within this 
model are shown in section \ref{collision} \cite{abspaper}. The elementary photoabsorption
cross section on the nucleon is discussed in section \ref{ele}.
Finally we present our results for the total photoabsorption
cross section on nuclei (section \ref{totcross}) \cite{abspaper} and the photoproduction of
pions (section \ref{pionprod}) and etas (section \ref{etaprod}) \cite{prodpaper}.
\section{The BUU model}
\label{model}
The BUU equation \cite{ber84,cass90} describes the classical time evolution of a many-particle
system under the influence of a self-consistent mean field potential and
a collision term. For the case of identical particles it is given by:
\begin{equation}
\label{buugl}
\frac{\partial f}{\partial t} + \frac{\vec{p}}{m}\, 
\frac{\partial f}{\partial \vec{r}}-
\vec{\nabla}U\, {\partial f \over \partial \vec{p}}=I_{coll}[f]
\quad,
\end{equation}
where $f(\vec{r},\vec{p},t)$ stands for the one-particle phasespace
density, $U[f]$ denotes the
self-consistent mean field potential and $I_{coll}[f]$ is the collision
term which - for a fermionic system - respects the Pauli principle.
For the description
of a system of non-identical particles one gets an equation for
each particle species that is coupled to all others by the
collision integral or the mean field potential.
Besides the nucleon we take all baryonic resonances up to
a mass of 2 GeV as well as the pion, the eta- and the rho-meson into account.
\par The collision term allows for the following reactions:
\begin{eqnarray*}
N\,N&\to&N\,N
\\
N\,N&\leftrightarrow& N\,R
\\
N\,N&\leftrightarrow& N\,N\,\pi \quad ({\rm S-wave})
\\
N\,R&\to&N\,R^{\prime}
\\
R&\leftrightarrow& N\,m
\\
R&\leftrightarrow& N\,\pi\,\pi
\\
&\leftrightarrow& \Delta(1232)\,\pi,\;N(1440)\,\pi,\;N\,\rho,\;
N\,\sigma
\\
\pi\,\pi&\leftrightarrow&\rho,\;\sigma \quad,
\end{eqnarray*}
where R stands for a baryonic resonance and m for a meson \cite{abspaper,teis}.
\section{Collision broadening}
\label{collision}
We have used the cross sections for the interaction of resonances
with nucleons from our transport model
to calculate the collision widths \cite{kondr94} of the resonances that are
important for photonuclear reactions \cite{abspaper}. These are the $P_{33}(1232)$, the $D_{13}(1520)$
and the $F_{15}(1680)$. The $S_{11}(1535)$ is important for the calculation of
etaproduction.
\par Since nucleon final states coming from spontaneous decays of resonances can
be Pauli blocked in the nuclear medium there is also a reduction of the
width. The total in-medium width is therefore:
\begin{equation}
\Gamma_{tot}^{med}=\Gamma_{spon}^{med}+\Gamma_{coll}^{med}\quad,
\end{equation}
where $\Gamma_{spon}^{med}$ stands for the sum of the one-pion-, two-pion- and
eta-width. 
\par
\begin{figure}[t]
\centerline{
\rotate[r]{\psfig{figure=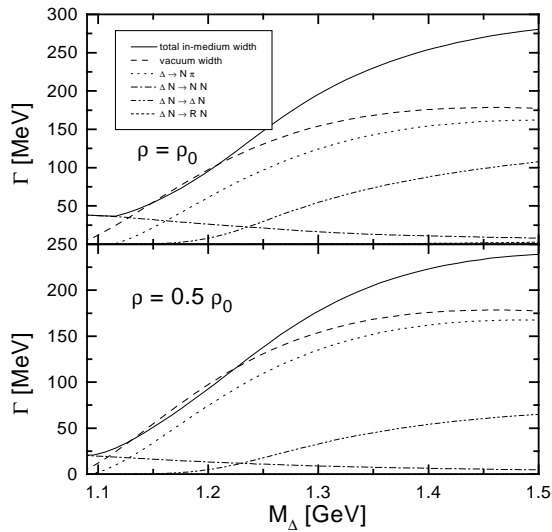,height=8cm}}}
\caption{In-medium width of the $\Delta$(1232).}
\label{delwidth}
\end{figure}

 In figure \ref{delwidth} we show the different contributions to the
in-medium width of the $\Delta(1232)$ at nucleon densities $\rho_0$ and
$\rho_0/2$ in isospin symmetric nuclear matter. Here the momentum of the
resonance is related to its mass by the requirement that the resonance was
created by photoabsorption on a free nucleon at rest. For comparison the
vacuum width is also shown.
\par At $\rho_0$ the collision width coming from
resonance absorption $N\,\Delta \to N\, N$ is at the pole of the resonance
($M$=1.232 GeV) about 25 MeV. This partial width decreases with increasing
mass. The contribution from $N\,\Delta \to N\,\Delta$ has - averaged over the
mass distribution - about the same size as the one from $N\,\Delta\to N\,N$.
However, here we get a strong increase of $\Gamma_{N \Delta \to N \Delta}$
with increasing mass because of the
phase space weighted integral over the mass distribution of the outgoing
$\Delta$-resonance. 
\par In the region of the resonance pole the total in-medium width is almost
independent of the nucleon density since collision broadening and Pauli
reduction of the free width nearly compensate. Thus, at the resonance pole
the net broadening
compared to the vacuum width is very small; at about 100 MeV above the pole
the width has grown by about 50 MeV, mainly due to the
$N \, \Delta \to N \, \Delta$ scattering process.
\par
\begin{figure}[t]
\centerline{
\psfig{figure=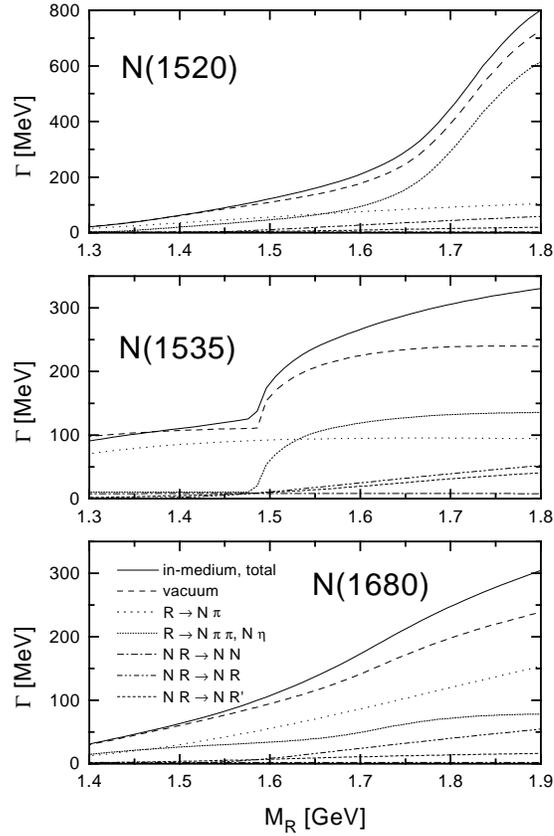,width=8cm}}
\caption{In-medium widths of the higher resonances that are relevant for
photonuclear processes.}
\label{hrestot}
\end{figure}

 In figure \ref{hrestot} the in-medium widths of the $N(1520)$, the
$N(1535)$ and the $N(1680)$ are compared with the vacuum widths and split
up into their partial widths. 
In all cases
the collision widths at the poles of
the resonances are only of the order 20 - 40 MeV which leads to a small net broadening
because the Pauli blocking of the free width is less important than in the
case of the $\Delta$-resonance. 
A collision width of 300 MeV \cite{alberico,kondr94} for the $D_{13}(1520)$ 
seems to be far from being realistic.
\section{The total photoabsorption cross section on the nucleon}
\label{ele}
For the one-pion production cross
section we use partial-wave amplitudes \cite{Arndt} and fit
the contributions coming from the $P_{33}(1232)$, $D_{13}(1520)$, $S_{11}(1535)$ 
and $F_{15}(1680)$ resonances to these amplitudes because - especially in
the region of the $\Delta$-resonance - interference terms with the background
are quite important \cite{prodpaper}. An incoherent de\-com\-po\-si\-tion of the total photoabsorption
cross section into resonance and background contributions as done by
Kondratyuk et al. \cite{kondr94} should not be used if one wants to
investigate possible modifications of the resonance contributions in nuclei. 
\par While the one-pion production cross sections can nicely be decomposed into
Breit-Wigner type resonance contributions and a smooth background, the
structure of the two-pion production cross sections is not described by the
resonance contributions that are induced by the two-pion decay widths of the
resonances \cite{prodpaper}. The difference between the
experimental cross section and the Breit-Wigner type resonance contributions
is treated
as background, where the momenta of the outgoing particles are distributed
according to three-body phase space. Therefore the only medium modification is
the possible Pauli blocking of the outgoing nucleon.
\subsection{Medium modifications}
We use the following medium modifications for the elementary photon-nuc\-le\-on
cross section:
\begin{itemize}
\item The vacuum width appearing in the resonance pro\-pa\-ga\-tors
is replaced by the in-medium width from
section \ref{collision}.
\item The collision width gives a Breit-Wigner type contribution to the
absorption cross section \cite{prodpaper}.
\item For the $\Delta$-resonance the difference between nucleon and
$\Delta$-potential causes a real part
of the self energy $\Pi$ to be used in the resonance pro\-pa\-ga\-tor:
\begin{equation}
{\rm Re}\,\Pi=2\,E_{\Delta}\,\left( U_N-U_{\Delta} \right) \quad.
\end{equation}
\item Nucleon final states can be Pauli blocked. 
\end{itemize}
\section{Results}
\subsection{Photoabsorption}
\label{totcross}
\par
\begin{figure}[t]
\centerline{
\psfig{figure=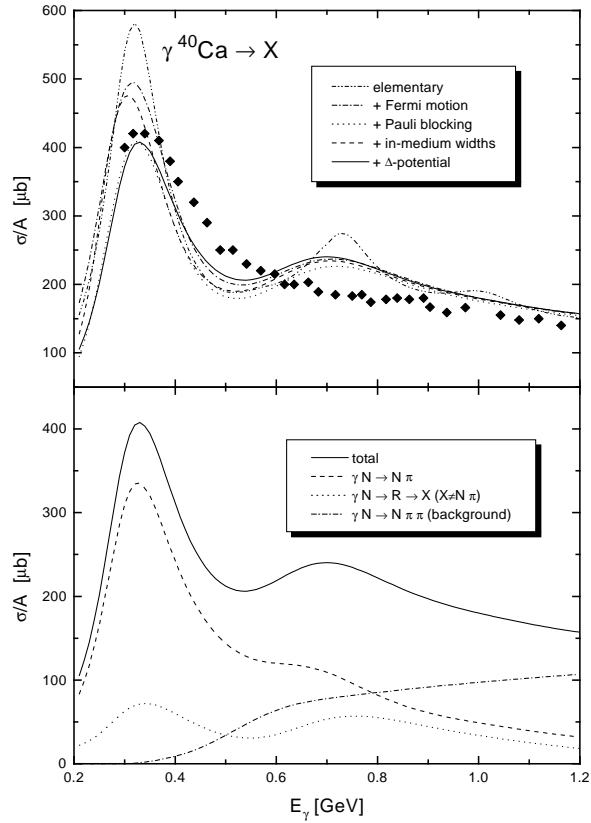,width=8cm}}
\caption{Photoabsorption cross section on $^{40}$Ca. The experimental data
are obtained by an average over different nuclei \protect\cite{bianchi}.}
\label{kgamabs}
\end{figure}

 Figure \ref{kgamabs} shows the total photonuclear cross sections on $^{40} {\rm Ca}$ that
result from successive application of the medium modifications. Fermi
motion alone leads to a damping of the $\Delta$-peak by about 100 $\mu$b per
nucleon. The structure in the region of the $D_{13}$-resonance is washed
out but does not disappear. The peak of the $F_{15}$-resonance vanishes.
Pauli blocking further decreases the $\Delta$-peak by about 100 $\mu$b with
the reduction of the cross section getting smaller at higher energies.
\par In these calculations the $\Delta$'s experience the same potential as the
nucleons. If we apply a $\Delta$-potential of $U_{\Delta}=-30 \,\rho/\rho_0\,
{\rm MeV}$ the $\Delta$-peak is
shifted to higher energies and decreased by about 70 $\mu$b per nucleon
because the $\Delta$-width increases strongly with increasing mass and the
$\Delta$-peak is proportional to $\frac{1}{\Gamma}$. Since the position
of the $\Delta$-peak becomes density dependent the integration over the
volume of the nucleus leads to a further smearing out. 
\par Compared to the
experimental data \cite{bianchi} we see from figure \ref{kgamabs} that we
underestimate the cross section in the high mass $\Delta$-region and 
overestimate the cross section in the region of the $D_{13}$-resonance. The
discrepancy in the $\Delta$-region can be resolved by inclusion of the two-body absorption process 
$\gamma \, N \, N \to \Delta \, N$ in our calculations \cite{prodpaper}. 
\par In figure \ref{kgamabs} (lower part) we also show the different contributions to the
total cross section. Here we see that the rise of the cross section between
550 and 700 MeV is not only caused by the excitation of the $D_{13}$-resonance
but also to a large extent by the opening of the two-pion background channel. The contribution
of the one-pion channel shows almost no resonant structure in this energy
regime.
\par We also calculated the photoabsorption cross section on $^{12}$C and
$^{208}$Pb. The peaks at the $\Delta$-resonance and at 700 MeV decrease 
slightly with increasing mass. The qualitative behaviour of the 
photoabsorption cross section does not depend on the mass number.
\par While the collisional broadening can, according to our results, not explain
the observed disappearance of the higher resonances, there are other
in-medium effects that still have to be explored. For example, there is
the possibility that the width of the $D_{13}$-resonance is
increased by a strong medium modification of the free width, for example
caused by the strong coupling to the $N\, \rho$-channel and a downward mass shift of
the $\rho$-meson in the nuclear medium \cite{rho}. This may lead to the disappearance
of the structure in the region of the $D_{13}$-resonance. Another
possibility is a strong medium modification of the elementary
$\gamma \, N \to N \, \pi \, \pi$ cross section in the nuclear medium or
a medium effect on the background amplitudes.
\par An understanding of the disappearance of the $D_{13}$-resonance in the
photoabsorption cross section might thus
be possible by a comparison between theory and experiment with respect
to more exclusive reaction channels.
\subsection{Pion production}
\label{pionprod}
\par
\begin{figure}[t]
\centerline{
\rotate[r]{\psfig{figure=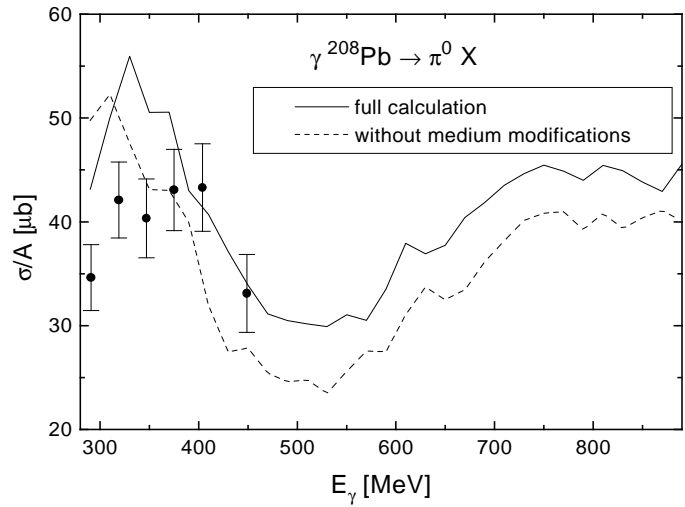,height=10cm}}}
\caption{Total $\pi^0$ photoproduction cross sections on
$^{208}$Pb. The experimental data are taken from \protect\cite{arends}.
The fluctuations in the calculated curves are caused by low statistics.}
\label{pbtpi}
\end{figure}

 The total $\pi^0$ cross section on $^{208}$Pb is shown in figure \ref{pbtpi} \cite{prodpaper}.
The curve labeled 'without medium modifications' results when applying only
Fermi motion and Pauli blocking and neglecting the difference between  
nucleon- and $\Delta$-potential and
the in-medium widths of the resonances. The difference to the full
calculation is basically caused by the $\Delta$-potential. In the
region of the $\Delta$-resonance the elementary absorption cross
section is reduced and shifted to higher energies \cite{abspaper}. For
larger photon energies the $\Delta$-potential leads to a slight
enhancement of the pion production due to its effect on the pion-nucleon
interaction.
Compared to the experimental data \cite{arends} our calculation
overestimates the cross section in the region of the $\Delta$-resonance
by about 20\% 
and we fail to explain the broad structure in the $\Delta$-region. The same holds
for calculations on $^{12}$C \cite{prodpaper}.  
\par The discrepancy in the $\Delta$-region might be due to a further reduction
of the cross section for $\gamma \,N \to N \,\pi$ in the nuclear medium.
A better description of the total photoabsorption cross section
then certainly requires 
the inclusion of two- and three-body absorption mechanisms
for the photon. This may also lead to a better description of the observed
structure of the cross sections.
\subsection{Eta production}
\label{etaprod}
We have parameterized the etaproduction on the free
nucleon under the assumption that the only production mechanism is
via an intermediate $N(1535)$-resonance \cite{prodpaper}. In nuclei we also have to take into
account eta production by final state interactions of pions that
were primary produced.
\par
\begin{figure}[t]
\centerline{
\rotate[r]{\psfig{figure=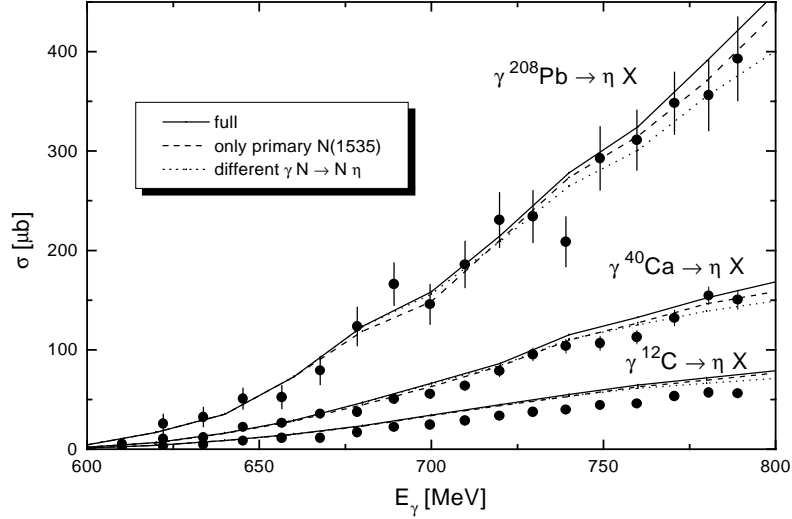,height=12cm}}}
\caption{Total eta photoproduction cross section on $^{12}$C,
$^{40}$Ca and $^{208}$Pb. All
experimental eta photoproduction data are taken from \protect\cite{robig-landau}.}
\label{calleta}
\end{figure}

 In figure \ref{calleta} we compare the calculated total etaproduction cross
section on $^{12}$C, $^{40}$Ca and $^{208}$Pb with experimental data \cite{robig-landau}. The
contributions coming from secondary processes are almost negligible.
For large photon energies our calculation depends on the choice of the
elementary $\gamma \,N \to N \,\eta$ cross section for photon energies
larger than 800 MeV because of the Fermi motion of the nucleons. An extrapolation
of this cross section according to low momentum transfer electroproduction \cite{prodpaper} 
reduces the total cross section at 800 MeV by about 15\% and
gives a better description of the experimental data.
On $^{12}$C there is only good agreement with the
experiment at low photon energies. For higher energies we overestimate
the cross section by about 20\%.
On $^{40}$Ca and $^{208}$Pb the agreement with the experiment is very good.
\section{Summary and outlook}
We have presented a calculation of the photoabsorption cross section and
cross sections for inclusive pion and eta photoproduction in
nuclei within a semi-classical BUU transport model
for photon energies
from 300 MeV to 1 GeV. Starting from a reasonable parameterization of the
free photon nucleon cross section we applied the medium modifications
Fermi motion, Pauli blocking and collision broadening for the involved
nucleon resonances.
\par For the $\Delta (1232)$-resonance
it turned out that collision broadening and reduction of the free width
by Pauli blocking nearly compensate each other resulting in a very small net
broadening. The collision broadening of the higher resonances in our
model is almost negligible.
\par Our calculated photoabsorption cross section fails to describe the
experimentally observed disappearance of the
$D_{13}$-resonance. This might be caused by a strong
broadening of the $D_{13}$-resonance in the nuclear medium due to a strong
coupling
to the $N \rho$-channel and a mass shift of the $\rho$-meson in the
nuclear medium
but also by
a medium modification of the $\gamma \,N \to N \, \pi \, \pi$ process.
\par In the $\Delta$-region we are able to reproduce the size of the observed
pion production cross sections in nuclei reasonably well 
although our calculated cross sections show too much structure in the
resonance region.
This might be due to the importance of multi-body
absorption mechanisms. 
\par The agreement of the calculated total eta production cross section with the
experiment is good. 
\par An experimental measurement of exclusive cross sections would be helpful for
a better understanding of the photon-nucleus reaction and an explanation
of the disappearance of the $D_{13}$-resonance in the total photonuclear
absorption cross section. In particular, the experimental investigation
of the 2$\pi$-channel on nuclei would be very important because of the
opening of this channel in the $N(1520)$-resonance region.

\section*{Acknowledgments}
This work was supported by BMBF, GSI Darmstadt and DFG.

\end{document}